\documentclass[aps,pra,10pt,twocolumn,groupedaddress,nofootinbib,notitlepage,showpacs,floatfix,longbibliography,superscriptaddress]{revtex4-1}

\usepackage{graphicx,graphics,epsfig,subfigure,times,bm,bbm,amssymb,amsmath,amsfonts,amsthm,mathrsfs,MnSymbol}
\usepackage[matrix,frame,arrow]{xypic}
\usepackage[pdfstartview=FitH]{hyperref}
\usepackage{subfigure}
\hypersetup{
    colorlinks=true,       
    linkcolor=red,          
   citecolor=magenta,        
    filecolor=magenta,      
    urlcolor=cyan,           
    runcolor=cyan
}
\usepackage[pdftex]{color}
\usepackage{braket}
\usepackage{enumerate}
\usepackage[normalem]{ulem}
\usepackage[usenames,dvipsnames]{xcolor}
\usepackage{multirow}
\usepackage{mathtools}

\definecolor{orange}{rgb}{1,0.5,0}

\newcommand{\bes} {\begin{subequations}}
\newcommand{\ees} {\end{subequations}}
\newcommand{\bea} {\begin{eqnarray}}
\newcommand{\eea} {\end{eqnarray}}

\definecolor{gold}{rgb}{0.85,.66,0}

\newcommand{\beq}{\begin{equation}}

\newcommand{\eeq}{\end{equation}}

\newcommand{\ignore}[1]{}
\newcommand{\mc}[1]{\mathcal{#1}}

\def\s{\sigma}

\def\>{\rangle}
\def\<{\langle}

\def\s0{I}

\newcommand{\ig}[1]{}

\begin{document}
\title{Exponential capacity of associative memories under quantum annealing recall}
\author{Siddhartha Santra}
\email{santra@stanford.edu}
\affiliation{U.S. Army Research Laboratory, Computational and Information Sciences
Directorate, ATTN: CIH-N, Aberdeen Proving Ground, Maryland, U.S.A. - 21005-5069.}
\affiliation{Department of Aeronautics and Astronautics, Stanford University, 496 Lomita Mall, Stanford, California, U.S.A. - 94305.}
\author{Omar Shehab}
\affiliation{Dept of Computer Science and Electrical Engineering, University of Maryland, Baltimore County, 1000 Hilltop circle, Maryland, U.S.A. - 21250.}
\author{Radhakrishnan Balu}
\email{radhakrishnan.balu.civ@mail.mil}
\affiliation{U.S. Army Research Laboratory, Computational and Information Sciences
Directorate, ATTN: CIH-N, Aberdeen Proving Ground, Maryland, U.S.A. - 21005-5069.}

\begin{abstract}
Associative memory models, in theoretical neuro- and computer sciences, can generally store a sublinear number of memories. We show that using quantum annealing for recall tasks endows associative memory models with exponential storage capacities. Theoretically, we obtain the radius of attractor basins, $R(N)$, and the capacity, $C(N)$, of such a scheme and their tradeoffs. Our calculations establish that for randomly chosen memories the capacity of a model using the Hebbian learning rule with recall via quantum annealing is exponential in the size of the problem, $C(N)=\mathcal{O}(e^{C_1N}),~C_1\geq0$, and succeeds on randomly chosen memory sets with a probability of $(1-e^{-C_2N}),~C_2\geq0$ with $C_1+C_2=(.5-f)^2/(1-f)$, where, $f=R(N)/N,~0\leq f\leq .5$ is the radius of attraction in terms of Hamming distance of an input probe from a stored memory as a fraction of the problem size. We demonstrate the application of this scheme on a programmable quantum annealing device - the Dwave processor.
\end{abstract}
\maketitle

\section{Introduction}
Associative memory models (AMM) are supervised learning models for the brain and reconstruct memories - desired configurations of quiescent and firing neurons - from input data that has only incomplete or incorrect information \cite{Hopfield}. To this end, the Hopfield network \cite{Hopfield,hopfieldscience} is a well established paradigm for associative memory where neurons are treated as binary threshold units \cite{mc-pitts} with their interconnections described by real weights. The network can be trained to memorise patterns - called \emph{learning} - via different algorithms \cite{learningrules} which evaluate the connection strengths based on the set of these fundamental memories. Once a network has learnt a certain number of patterns it should \emph{recall} an initial (possibly imperfect) bit string configuration to a stored pattern which had maximum overlap with the initial state, and this is interpreted as successful recognition. Each distinct combination of a learning rule and recall method corresponds to a different associative memory model. Even in the canonical setting for Hopfield networks, where the neurons are considered to be classical Ising spins, succesful memory recall amounts to global minimization of an cost (energy) function over the possible collective spin configurations \cite{nnspinglass,hopfieldscience,talagrandbook}. The classical dynamical update rules however do not guarantee that this global minimum is indeed reached - often the asymptotic state is a local energy minimum \cite{spurious}. While AMMs trace their origins to theoretical neuroscience, they have been widely considered in the classical setting, and an area of active current research in the quantum setting, in the context of  content-addressable memories \cite{cams}, machine learning \cite{qml-intro,qml-advances}, artificial neural networks \cite{q-ai,denchev} and neuromorphic computing \cite{neural-comp}. 

In this work we demonstrate the use of quantum annealing (QA) \cite{physqa,qa-revmodphys} for recalling stored memories in AMMs. QA is a non-universal form of adiabatic quantum computation (AQC) \cite{farhiaqc} that solves hard optimization problems \cite{santramax2sat} by encoding the solution into the lowest energy state of a problem-defined Hamiltonian. The search for the global energy minimum is assisted via quantum tunneling across barriers in the energy landscape which reduces the chances of getting trapped in local minima \cite{mathqa}. Leveraging the robustness \cite{childs-robustaqc,sarandy2005} of open-system AQC, QA has become a promising scalable quantum sub-routine for the solution of practical computational problems \cite{santramax2sat,Zick2015} using currently available technology \cite{dwave-tech}.

By casting the problem of succesful memory recall in associative memory models as one of finding the spin configuration corresponding to the global energy minimum under a Hamiltonian, determined in part by the stored memories and partially by the imperfect input memory, one can hope to use QA for recall tasks in AMM. Just as in the classical AMM case, the memories are encoded as the weights of a fully connected network of spins - qubits in our case. However, in contrast to the classical case, the probe memories are input to the system as local field biases on the qubits and not as their initial values. We show that using QA for recall tasks \cite{neigovzen:2008,seddiqi:2014} in associative memory models, with any learning rule that does not discriminate against any of the fundamental stored memories, leads to an exponential storage capacity for randomly chosen memories and succeeds, over random choices of fundamental memory sets, with a probability approaching unity exponentially in the problem size. We also demonstrate an implementation of quantum annealing recall in an associative memory model with the Hebbian learning rule on a programmable quantum annealing device - the Dwave processor \cite{dwave-tech,QAD}. Our results are valid for completey general Hopfield networks and may be contrasted with purely classical schemes for AMMs that require special pattern classes or connectivity structures \cite{class-exp-mem} in order to achieve super-polynomial capacity.

The paper has three following sections with the theoretical setup described in Sec.~(\ref{thsetup}), implementation results with the Dwave processor in Sec.~(\ref{exptlsec}) and concludes with a discussion in Sec.~(\ref{discconc}).
\section{Theoretical Setup}
\label{thsetup}
Here, the framework of Hopfield networks is first discussed in subsec.~(\ref{subsec:hnam}) along with its use in AMMs. We then describe the process of quantum annealing from the perspective of finding ground states of classical Hamiltonians  in subsec.~(\ref{subsec:qa}). Followed by a discussion of how quantum annealing may be used for recall tasks in AMM, subsec.~(\ref{subsec:qar-amm}). As with any update rule in the classical setting only those input probe memories which lie within a certain maximum Hamming distance from any of the stored fundamental memories - the radius of attraction - may be successfully recalled using QAR-AMM which is discussed in subsec.~(\ref{subsec:rofa}). Finally, we obtain the capacity of the total scheme where the learning is done via the Hebbian rule with quantum annealing recall in subsec.~(\ref{subsec:cap}) and show the tradeoff between the capacity and the size of the radius of attraction. In the same section we discuss the probability of our scheme succeeding over randomly chosen fundamental memory sets.

\subsection{Hopfield network and Associative memory}
\label{subsec:hnam}
The Hopfield network (HN) is a fully connected graph $K_N$ of interacting binary state `neurons' $\{S_i=\pm1\}_{i=1,2...,N}$ whose weights, $W_{ij}$, encode the bit strings (of size $N$) that form its memory $M:=\{\xi^\mu\}_{\mu=1,...,p}$. Given the set of memories, $M$, the use of different learning rules lead to different entries $W_{ij}$ for the weight matrix. In this paper we consider the Hebbian learning rule whose weights $W_{ij}$ for $i\neq j$ are given by,
\begin{align}
W_{ij}&:=\frac{1}{N}(\sum_{\mu=1}^p\xi^\mu_i\xi^\mu_j-p\delta_{ij})~\forall i,j\in[1,N]
\label{hebbrule}
\end{align}

The Hebbian learning rule \cite{hebbrule} has the characteristics of being local, incremental and immediate. Locality here means that a connection weight depends only on the state of the spins across the connection. Incrementality implies that new memories can be learnt without referring to previously learnt memories and immediate means that the connection weights for any number of memories can be obtained in a finite number of steps. The absolute storage capacitiy $C(N)$ defined as the number of memories one can store in a network of size $N$ when perfect recall accuracy is desired is $C(N)=N/2\log(N)$ for the Hebbian rule \cite{hopcapacity}.

In conjunction with a learning rule a Hopfield network may be used as an associative memory model i.e. as a content addressable memory where given an initial configuration of neurons $S^I\in \{0,1\}^N$ (considered the input or probe memory vector) the dynamics of the network ideally results in a final network configuration $S^F$ which is some  stored memory (also called fundamental memory) $\xi\in M$ that was closest in Hamming distance from the original configuration $S^I$. This dynamic is implemented as an update rule for the spins,
\begin{align}
S_i(t+1)\to\text{Sign}(\sum_{j}W_{ij}S_j(t)-\theta_i)
\label{classicalupdate}
\end{align}
where $S_i(t+1)$ is the value of spin $i$ in the timestep after $t$. One can understand the attractor dynamics as a search for a global energy minimum by attaching an energy value to a configuration of spins $S=(S_1,S_2,...,S_N)$ in the network through the definition,
\begin{align}
E(\bar{S}):=-\sum_{i<j}W_{ij}S_iS_j-\sum_i\theta_iS_i,
\label{eq:classen}
\end{align}
where $\theta_i$ is the threshold value for spin $i$ in the network. The Markovian dynamics generated by the rule (\ref{classicalupdate}) ensures that $E(\bar{S})$ is non-increasing during the evolution. The asymptotic fixed point is thus (at least) a local minimum and the corresponding spin configuration is a stable attractor for the dynamics. The local threshold values $\theta_i$ can serve to bias (or even freeze) certain spin values to the ones desired.

\subsection{Quantum annealing}
\label{subsec:qa}
Quantum annealing (QA) is a finite temperature, non-universal form of Adiabatic Quantum computation (AQC) useful for solving hard optimization problems. Given the cost function, $Cost(X):\{0,1\}^n\to\mathbb{R}$, of an optimization problem, QA finds the configuration, $X$, a vector of Boolean variables obtained after a qubit-wise read out of a quantum state - that minimizes $Cost(X)$. Thus QA can also be understood as the quantum couterpart of simulated annealing \cite{kirkpatrick1983}. In general, the cost function can be encoded as a Hamiltonian operator $\hat{H}_P$ whose ground state encodes the solution to the computational problem. Of interest in QA is the final ground state of the time dependent Hamiltonian,
\begin{align}
\hat{H}(t)=A(t)\hat{H}_I+B(t)\hat{H}_P,
\label{eq:timedepham}
\end{align}
which undergoes \emph{annealing} as the classical control process takes the parameters from (effectively), $A(t=0)=1,B(t=0)=0$ to $A(t=T_{\text{anneal}})=0,B(t=T_{\text{anneal}})=1$, where, $T_{\text{anneal}}$ is the duration of the annealing process. The expectation is that the ground state of the initial Hamiltonian $\hat{H}_I$ is easily-prepared and annealing takes it to the ground state of the final Hamiltonian $\hat{H}_P$ which can also be read out easily to yield the solution. QA requires that the Hamiltonian $\hat{H}_P$ be diagonal in the computational basis which means that the process of reading out the final state does not introduce any further complexity to the computational problem beyond requirements of adiabaticity \cite{farhiaqc,sarandy2005} during the anneal process. In practice, for the specific QA device we use, Sec.~(\ref{exptlsec}), QA at non-zero temperature $T$ starts in the limit of strong transverse field terms in $\hat{H}_I$ and weak $\hat{H}_P$, i.e. $A(0)\gg \text{max}\{k_BT,B(0)\}$, with the initial ground state close to an equal superposition of all computational basis states of the qubits in the problem. Monotonically decreasing $A(t)$ and increasing $B(t)$ takes the system (close) to the ground state of $\hat{H}_P$ as at the final time $B(T_{\text{anneal}})\gg A(T_{\text{anneal}})$ \cite{QAD}.

 The QA process relies on the quantum adiabatic theorem - thus the annealing duration $T_{\text{anneal}}$ has to be sufficiently large and the temperature $T$ of the system sufficiently high to prevent diabatic transitions away from the instantaneous ground state of $H(t)$. For any given problem, i.e. $H_P$, the initial Hamiltonian $H_I$ and the temporal dependence of $A(t),B(t)$ - the minimum required values for $T_{\text{anneal}}$ and the maximum allowed value of $T$ are determined by the inverse energy gap between the instantaneous ground state and the first excited state of $H(t)$ at any $t\in[0,T_{\text{anneal}}]$.

\subsection{Recall tasks in AMM using Quantum Annealing}
\label{subsec:qar-amm}
The energy function (\ref{eq:classen}) of a HN admits a physical interpretation as the Hamiltonian operator of a spin-glass problem (with local field terms) \cite{ising-spinglass} where the binary-state neurons may be treated as Ising spins interacting with each other via their couplings $J_{ij}=W_{ij}$ and the threshold values may be interpreted as local fields $h_i=\theta_i$. By making the correspondence $S_i\mapsto \hat{\sigma}_i^z$ the energy function (\ref{eq:classen}) may be identified with the Hamiltonian operator, $\hat{H}_{\text{AM}}=-\sum_{i>j}J_{ij}\hat{\sigma}^z_i\hat{\sigma}^z_j+\sum_{i}h_i\hat{\sigma}^z_i$,  for the system whose set of ground states (the set containing all the degenerate lowest energy states) encodes the configurations the network \emph{has committed to its memory}. If an eigenvector $\ket{\chi}$ of $\hat{H}$ is an element of the set of ground states then the corresponding bit string $\chi$ with entries $\chi_i=\bra{\chi}\hat{\sigma}^z_i\ket{\chi}$, is a memory state only if $\chi$ corresponds to one of the stored memories, i.e. $\chi\in M$, otherwise it is a spurious state \cite{spurious,hopfieldscience} and corresponds to incorrect memory recall. Since the memories are all encoded only in the coupling terms between different spins we call  $\hat{H}_{\text{mem}}:=-\sum_{i>j}J_{ij}\hat{\sigma}^z_i\hat{\sigma}^z_j$ the memory Hamiltonian whereas, as we show below, the local field term can be used to probe memory recall using an input state and hence is called the probe term i.e. $\hat{H}_{\text{probe}}:=\sum_{i}h_i\hat{\sigma}^z_i$. We use the Hamiltonian $\hat{H}_{\text{AM}}$ as the final problem Hamiltonian in Eq.~(\ref{eq:timedepham}), i.e. $\hat{H}_P=\hat{H}_{\text{AM}}$, with the probe memory part dependent on an input string $\chi$, of length $n:=|\chi|$, $0\leq n \leq N$, given by $\hat{H}_{\text{probe}}=-h\sum_i\chi_i\hat{\sigma}^z_i$ where $h>0$ is some overall scale. Thus,
\begin{align}
\hat{H}_P=\hat{H}_{\text{AM}}=-\sum_{i>j}J_{ij}\hat{\sigma}^z_i\hat{\sigma}^z_j-h\sum_{i}\chi_i\hat{\sigma}^z_i
\label{eq:ham}
\end{align}

While the memory hamiltonian $\hat{H}_{\text{mem}}$ is degenerate for all stored memories $\ket{\xi^\mu}\in\mc{M}$, the probe Hamiltonian $\hat{H}_{\text{probe}}$ breaks this degeneracy as follows,
\begin{align}
\hat{H}_{\text{probe}}\ket{\xi^\mu}&=-h(\sum_{i|\chi_i=\xi^\mu_i}1-\sum_{i|\chi_i=-\xi^\mu_i}1)\ket{\xi^\mu}\nonumber\\
&=-h(\sum_{i=1}^{i=|\bar{\chi}|}1-2\sum_{i|\chi_i=-\xi^\mu_i}1)\ket{\xi^\mu}\nonumber\\
&=-h(n-2d^\mu_\chi)\ket{\xi^\mu}
\label{eq:probemem}
\end{align}

where $n$ is the length of the input probe bit string and $0\leq d^\mu_\chi\leq n$ is the Hamming distance between the input bit string $\chi$ and the bit string $\xi^\mu$ corresponding to the memory eigenvector $\ket{\xi^\mu}$. Eq.~(\ref{eq:probemem}) implies that the probe Hamiltonian energetically orders the stored memories according to their Hamming distances from the input bit string.

We thus have a scheme for a quantum annealing implementation of an AMM using Eqs.~(\ref{eq:timedepham}) and (\ref{eq:ham}). Starting from the ground state of an arbitrary initial Hamiltonian $\hat{H}_I$ (generally fixed by experimental limitations) if one can tune the temporal evolution of $\hat{H}(t)$ to arrive at the final Hamiltonian $\hat{H}_P=\hat{H}_{\text{AM}}$ while maintaining conditions of adiabaticity then the final ground state should be the memory we hoped to recover. We call this scheme quantum annealing recall in associative memory models (QAR-AMM).

The requirement that the energy,  relative to the Hamiltonian (\ref{eq:ham}), for any input probe vector that is not one of the fundamental memories be greater than for any of the fundamental memories leads to a bound on the maximum value of the field strength $h$ for successful recall, $
E(\ket{\xi^\mu})<E(\ket{\chi})\implies h<(\braket{\xi^\mu|W/2|\xi^\mu}-\braket{\chi|W/2|\chi})/2d^\mu_\chi$. This maximum field strength can be shown to depend on the number of random memories stored and the Hamming distances of the input probe memory from the stored fundamental memories in general, however when the fundamental memories are mutually orthogonal to each other i.e. every pairwise Hamming distances is $N/2$, this maximum value evaluates to, see Appendix.~(\ref{hboundmax}),
\begin{align}
h<\frac{1}{4d^\mu_\chi}[N(1-p)+4\sum_{\nu=1}^pd^{\nu}_\chi-\frac{4}{N}\sum_{\nu=1}^p(d^\nu_\chi)^2]=:h^\mu_{\chi,\text{max}}
\label{hbound}
\end{align}
If we restrict ourselves to working with positive field biases then we get that for succesfully recalling a memory the allowed values of $h$ are $0<h<\text{max}_{\mu}~h^\mu_{\chi,\text{max}}$.

\subsection{Radius of attraction using QAR-AMM.}
\label{subsec:rofa}
The memory Hamiltonian, $\hat{H}_{\text{mem}}$, part of $\hat{H}_{\text{AM}}$ has a global spin flip symmetry which implies that for each memory eigenket $\ket{\xi^\mu}\in\mc{M}$ the spin-flipped state $\ket{\tilde{\xi}^\mu}=\otimes_{i=1}^N\hat{\sigma}^x_i\ket{\xi^\mu}$ - which would be a spurious state for purposes of memory recall - is also degenerate with respect to $\hat{H}_{\text{mem}}$. While the probe Hamiltonian breaks the degeneracy of memory states $\ket{\xi^\mu}\in\mc{M}$ in the desired manner, we find that it also shifts the energy of these spurious spin-flipped states in the reverse manner,
\begin{align}
\hat{H}_{\text{probe}}\ket{\tilde{\xi}^\mu}&=h(n-2d^\mu_\chi)\ket{\tilde{\xi}^\mu},
\end{align}

i.e., spurious states that are \emph{further} from the input memory in Hamming distance have lower energies w.r.t. $\hat{H}_{\text{probe}}$. This means that for a set of $p$ memories if a given input state $\ket{\chi}$ is nearest to $\ket{\xi^\mu}$ and furthest from $\ket{\xi^\nu}$ at Hamming distances $d_\chi^\mu,d_\chi^\nu$ respectively (we call the shortest distance $d^s_\chi:=\text{min}_{\mu}d^\mu_\chi$ and the largest $d^b_\chi:=\text{max}_{\mu}d^\mu_\chi$) then by requiring that the energy w.r.t. $\hat{H}_{\text{probe}}$ of the nearest memory state be lower than that of the lowest energy spurious state we arrive at the condition,
\begin{align}
d^s_\chi+d^b_\chi&\leq (n-1)=(|\chi|-1),
\label{ineq:hamdistbound1}
\end{align}
which along with the definition $d^s_\chi\leq d^b_\chi$ results in $d^s_\chi<\lfloor n/2\rfloor$. Thus all full length ($|\chi|=N$) input states $\chi$ within a radius,
\begin{align}
R(N)&\leq (N-2)/2~~~\text{N: Even}\nonumber\\
&\leq (N-1)/2~~~\text{N: Odd}
\label{radius}
\end{align}
are attracted to the closest fundamental memory and thus defines its basin of attraction.

Further, using the triangle inequality for Hamming distances we can also lower bound the sum $d^s_\chi+d^b_\chi$ by the maximal Hamming distance between any two vectors in the memory set for the given combination of $n=|\chi|$ bits. That is given 
$d(n):=\text{max}_{\mu,\nu}~d_n(\bar{\xi}^\mu,\bar{\xi}^\nu)$ where $d_n$ is the Hamming distance between $\bar{\xi}^\mu,\bar{\xi}^\nu$ for a particular combination of $n$ bits - we have in combination with Ineq.~(\ref{ineq:hamdistbound1}) \footnote{Since Hamming distances can only take natural number values, the strict Ineq.~(\ref{ineq:hamdistbound1}) means $d^s_\chi+d^b_\chi$ can atmost equal $(n-1)$.},
\begin{align}
d(n)\leq d^s_\chi+d^b_\chi&\leq (n-1).
\label{ineq:hamdistbound2}
\end{align}
which describes the set of all input vectors $\bar{\chi}$ that can be successfully recalled using QAR-AMM. Clearly for smaller $d(n)$ values there are a larger number of states within the attraction basin. Intuitively, this implies that for set $M$ of memory vectors with a smaller span in Hamming distances, but well separated within this span, the QA-AMM scheme works well. On the other hand for $n=N$ if, for example, $d(N)=N$, meaning that both a bit string and its negation are in the memory set, then the scheme fails.

\subsection{Capacity, Attraction Basin size and tradeoffs}
\label{subsec:cap}
The capacity $C(N)$ of a model for associative memory for bit strings of length $N$ is defined as the number of randomly chosen bit strings that may be stored in the network with the requirement that these states be stable fixed points under the dynamics dictated by the update rule for the spins. Thus in the classical setting the capacity depends on the learning rule as well as the update rule for network dynamics. This is true using the QAR-AMM for memory recall as well. 

In our scheme, when the memories are randomly chosen in a balanced manner, i.e. the probability of a bit being $1(-1)$ in any of the $p$ memories is $.5(.5)$, there is a finite probability that an input probe memory $\chi$ which even though is within the basin of attraction of a fundamental memory fails to be recalled correctly because its distance from the furthest memory, $d^b_{\chi}$, violates inequality (\ref{ineq:hamdistbound1}) i.e. $d^b_{\chi}>N-1-d^s_\chi$. The probability of this happening can be made to approach zero exponentially in the size $N$ provided the radius of attraction (\ref{radius}) is reduced by a constant fraction of $N$ from the one given in Eq.~(\ref{radius}). Allowing for this non-zero failure probability at any finite $N$ - the capacity turns out to be exponential in the size of the problem.

To calculate the capacity we consider a set of $p$ fundamental memory vectors, $\xi^{\mu}$, of length $N$ (even here for ease of presentation) whose $pN$ entries are discrete random variables, $(\xi^{\mu})_i=\pm1$, that are i.i.d. with equal probability $=.5$. Suppose now we consider an input probe memory $\chi$ at a hamming distance $d^s_\chi=(N-2)/2-x,~x=0,1,...,(N-2)/2$ from some memory that we call $\xi^1$. Then using Ineq.~(\ref{ineq:hamdistbound1}) QAR-AMM succeeds if any \emph{other} fundamental memory vector is at a Hamming distance of atmost $d^b_\chi\leq N/2+x$. The probability that this happens for any one other fundamental memory $\xi^\mu,~\mu\neq 1$ is given by \cite{lovaszbook},
\begin{align}
P[d^\mu_\chi\leq (N/2+x)]&=\sum_{l=0}^{(N/2+x)}P(d^\mu_\chi=l)=\sum_{l=0}^{(N/2+x)}\frac{\binom{N}{l}}{2^N}\nonumber\\
&\geq 1-\frac{1}{2}\text{exp}(\frac{-x^2}{N/2+x})\nonumber\\
&=1-.5\text{exp}(\frac{-t^2}{.5+t}N)=:P^*.
\label{p*}
\end{align}
where $x=tN,~0\leq t<.5$. Intuitively this means that as the radius of the attraction basin is allowed to decrease by Hamming size $x=tN$, the probability of the scheme succeeding when there are only two stored memories approaches unity exponentially in $N$.

Since the memories are independently chosen - the probability that all the $(p-1)$ memories apart from the one which has Hamming distance $d^s_\chi$ from $\bar{\chi}$ have distances $d^\mu_\chi\leq (N/2+x)~\forall \mu$ is lower bounded by $(P^*)^{p-1}$. If we now require that having stored $p$ fundamental memories our scheme succeeds with probability at least $\gamma$, i.e. $(P^*)^{p-1}\geq \gamma$, then taking the logarithm of both sides we obtain a bound on the number of memories the network can possibly store,
\begin{align}
p\leq (1+\frac{\log \gamma}{\log P^*})
\label{cap1}
\end{align}

We demand that asymptotically in $N$ we get perfect recall and require that this approach be exponential. Then we have that,
\begin{align}
\gamma=(1-e^{-C_2N}),~C_2>0
\label{gamma}
\end{align}

Then, using a small $z$ approximation for $\log(1-z)\simeq -z$ and Eqs.~(\ref{p*},\ref{gamma}) in Ineq.~(\ref{cap1}) we get,
\begin{align}
p&\leq 1+2~\text{exp}(-C_2N+\frac{t^2}{.5+t}N)\nonumber\\
&=1+2~\text{exp}(C_1N)=O(e^{C_1N}),
\label{cap2}
\end{align}
where $C_1=t^2/(.5+t)-C_2$. Eq.~(\ref{cap2}) thus implies an exponential capacity for $0\leq C_2\leq t^2/(.5+t)$.

There is also a tradeoff between the size of the attraction basin and the capacity, just as in the classical setting \cite{storkeyffl}, which can be seen by obtaining the relationship between the constants $C_1,C_2$ in terms of the radius of the basins of attraction, $f=R(N)/N=(N/2-1-x)/N=(N/2-1-tN)/N\to_{N\to\infty}(.5-t),~0\leq f<.5$, resulting in,
\begin{align}
C_1+C_2=\frac{(.5-f)^2}{(1-f)}.
\label{c1c2}
\end{align}

Note that ideally one would want both $C_1,C_2$ to be as large as possible because, respectively, they represent the exponent for the exponential capacity and the probability for QAR-AMM to succeed. However, the R.H.S. of Eq.~(\ref{c1c2}) is a positive valued monotonically decreasing function of $f$. Thus smaller values of $f$ would imply higher capacity and scheme success probability but smaller radii of attraction basins and vice versa.

\section{Quantum annealing recall with a programmable quantum annealer.}
\label{exptlsec}
In this section we present results from experimental implementations of the QAR-AMM on the Dwave quantum annealing processor. In Subsec.~(\ref{subsec:expsetup}) we describe the Dwave quantum annealing processor and the settings we use, a description of the required embedding of our fully connected networks onto the native qubit connectivity on the processor chip in subsec.~(\ref{subsec:embed}) and finally examples of memory recall using quantum annealing in subsec.~(\ref{subsec:example}).

\subsection{Experimental Setup}
\label{subsec:expsetup}

\begin{figure}
\centering
    \includegraphics[width=.9\columnwidth,height=7cm]{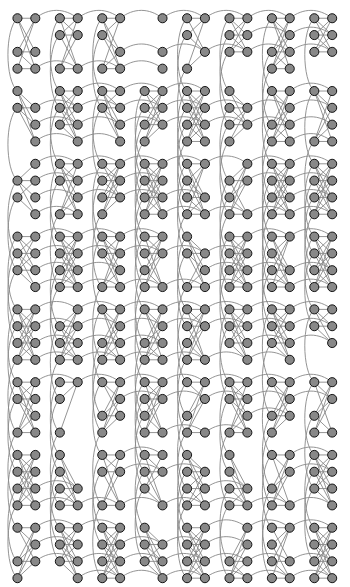}
    \caption{`Chimera' graph showing the connectivity of qubits on the DW2 processor chip at Burnaby, BC that we use. Not all qubits are usable in the graph - missing qubits - which are rejected at the calibration stage. There are 64, $K_{4,4}$-connected blocks of qubits laid out as a matrix of $8\times 8$ blocks. Each block has 2 columns (vertical) and 4 rows (horizontal). Fully connected problems such as Hopfield networks have to be embedded onto the native graph structure keeping in mind the missing qubits.}
    \label{fig:dw2}
\end{figure}

To demonstrate associative memory recall using quantum annealing we use the second generation of the commercially available Dwave processors \cite{dwave-tech}, DW2, with \ensuremath{512} qubits of which \ensuremath{476} qubits are effectively available. These qubits form the nodes of the so-called ``Chimera" graph shown in Fig.~(\ref{fig:dw2}). The engineering and physics of the processor chip has been extensively discussed in the literature \cite{dwhardware1,dwhardware2} and references therein. In this paragraph we briefly touch upon some features that are relevant to our problem. The DW2 chip comprises of superconducting
rf SQUID flux qubits interacting with each other via Josephson junctions and is maintained at a base temperature of $T\simeq 15 mK$ in a dilution refrigerator. A classical program supplies the problem Hamiltonian to the chip via the $N\times N$ matrix of coupling values, $J_{ij}$, and a $N\times 1$ vector of field strength, $h_i$, values. These final $J_{ij},h_i$ values are achieved on the processor at the end of the user set annealing time $T_{\text{anneal}}$ which can be set at integer values between $20~\mu s$ to $20,000~\mu s$ with the default value being $20~\mu s$. The time to wait after this programming, called thermalization time, in order for it to cool back to base temperature can be between \ensuremath{0} to \ensuremath{10000} microseconds with the default value being \ensuremath{10000} microseconds. The coupler strengths $J_{ij}$ can be set between $J_{ij}\in[-1,1]$ while the local fields between $h_i\in[-2,2]$. These values go through a non-linear 9-bit analog to digital conversion (ADC) and thus the step size for either is the extent of their range divided by $2^9$ - so the $|J_{ij}|$ values can be set in multiples of $2^{-8}$ while $|h_i|$ in steps of $2^{-7}$, however there are noise contributions which are important at low values of $h,J$ \cite{privcommthom}. The time to wait after each state is read from the processor in order for it to cool back to base temperature is $0 \mu s$ as the readout process is not supposed to supply any thermal noise. 

For our experiments - to minimize diabatic transitions due to finite annealing times we use the maximum allowed annealing time $T_{\text{anneal}}=20,000~\mu s$ and the maximum allowed thermalization time of $10,000~\mu s$. Further, we choose a problem defined on $N=2^4$ qubits so that the values of $|J_{ij}|=O(2^{-4})$, given by the Hebbian weight matrix Eq.~(\ref{hebbrule}), are much larger compared to its resolution.

\subsection{Embedding fully connected Hopfield networks in Chimera}
\label{subsec:embed}

A major step in solving a problem on the Dwave Two computer is mapping a generic Ising problem Hamiltonian, such as $\hat{H}_{\text{AM}}$, to the Dwave's native chimera graph which is a composition of \ensuremath{K_{4,4}} graphs (complete bipartite graph with $4$ vertices in each partition) - an instance of the minor embedding problem which is {\bf NP}-complete \cite{garey2002computers}. A problem can be embedded in more than one way. The  Dwave API provides the function `\texttt{find\_embedding(J)}' that uses a heuristic algorithm to perform the embedding which works reliably when the number of logical qubits, $N$, is under \ensuremath{50}. The input argument to function is the $N\times N$ coupling matrix, $J$, and the algorithm looks at the adjacency matrix derived from, $J$, to obtain a possible embedding \cite{minor-embed}. Asymptotically, roughly $N$ completely-connected logical qubits may be embedded on a hardware chimera graph of \ensuremath{N^2} physical qubits. On the particular \ensuremath{512}-qubit DW2 processor we use, only \ensuremath{476} qubits are available to be programmed as shown in Fig.~(\ref{fig:dw2}), the remaining qubits being unusable due to hardware faults. We note that not all problems require complete graphs hence larger non-trivial problem graphs can be embedded depending on which problem is being attempted. For example, the problem graph for the graph isomorphism problem on a Dwave machine is not a complete graph \cite{Zick2015} nor are certain restrictions \cite{class-exp-mem} to the canonical Hopfield networks.

When a problem is embedded on a hardware graph, a logical qubit is represented by a ferromagnetic chain of physical qubits. Ideally, after annealing, all the physical qubits are in the same state carrying the value of the state of the logical qubit. In reality, the chain tends to break down more often when it becomes longer, i.e., some physical spins corresponding to the same logical variable do not agree. In this scenario, one needs to use gauge averaging \cite{QAD}, majority voting \cite{Zick2015} or the more general quantum error-correcting schemes \cite{pudenz2015quantum}.

For the representative example discussed in the next subsection we have not implemented any error-correction strategy. This lets us discuss the raw implementation of QAR-AMM with respect to the theory in the previous section.

\subsection{Representative example}
\label{subsec:example}

\begin{figure}
\centering
    \includegraphics[width=.9\columnwidth]{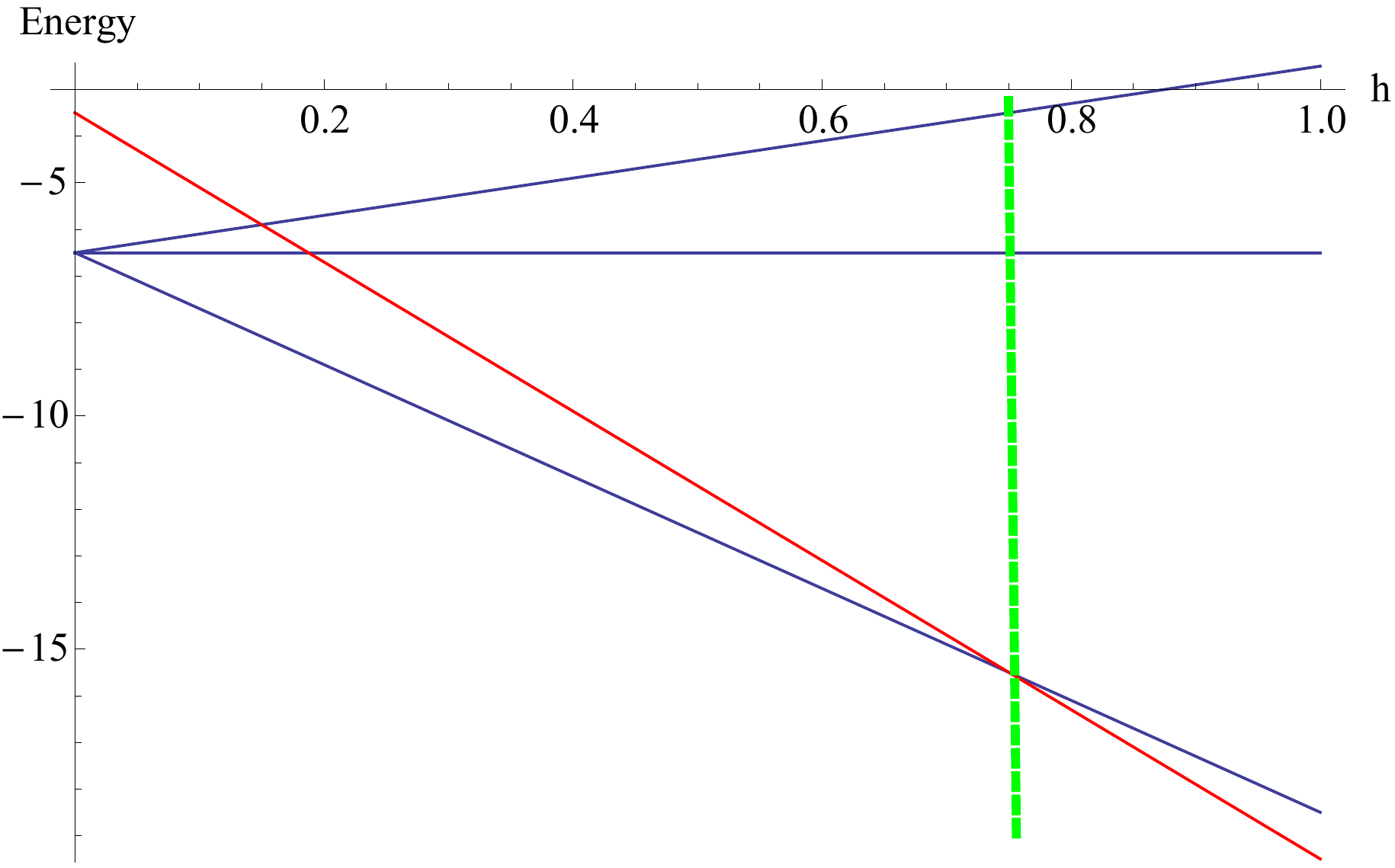}
    \caption{(color online) The variation of the energies of the fundamental memories and the probe memory under the Hamiltonian $\hat{H}_{\text{AM}}$ with the probe vector $\bar{\chi}$ as in (\ref{probe1}). The dotted vertical line (green) represents the highest ($h=.75$) allowed field strength for succesful recall of $\bar{\chi}$. Applying fields above this maximum value overbiases the Hamiltonian such that $\bar{\chi}$ itself becomes the lowest energy state. A vertical slice at any fixed value of $h$ is the spectrum of the problem Hamiltonian $\hat{H}_P=\hat{H}_{\text{AM}}$ for the $p$-fundamental memories plus the input probe memory.}
    \label{fig:example-energy}
\end{figure}

\begin{figure}
\centering
    \includegraphics[width=\columnwidth]{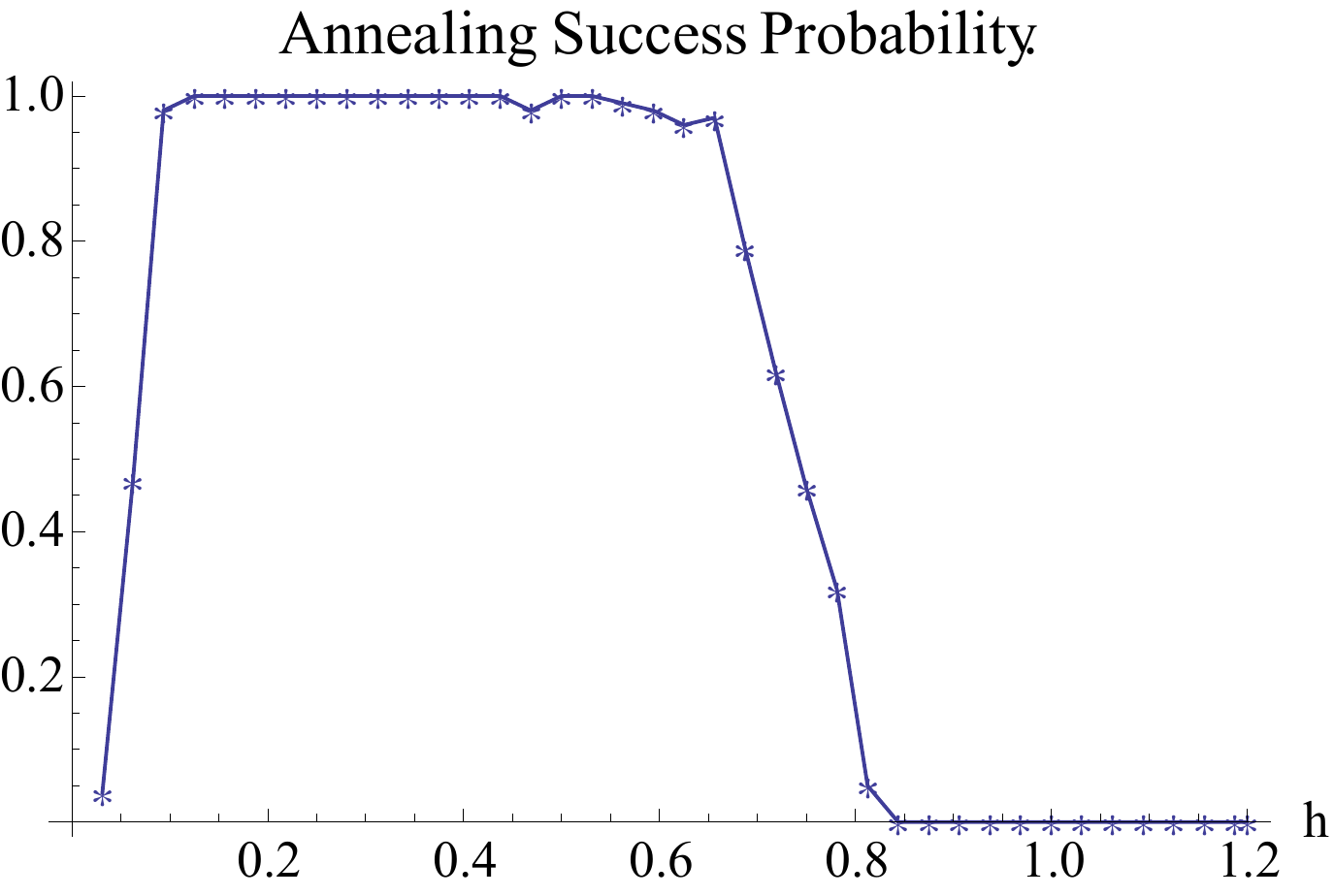}
    \caption{(color online) Probability of the correct recall using quantum annealing varying with respect to the applied field strength $h>0$. This probability is high ($\simeq 1$) for the particular set of $p=3$ memories and the input vector (\ref{probe1}) for almost the entire region with $h<.75$ (green dashed line). For small values of $h$ ($\leq .15$), the thermal noise degrades the annealing recall success significantly.}
    \label{fig:example}
\end{figure}

We demonstrate the actual implementation of the QAR-AMM scheme using the Dwave quantum annealing hardware by describing a representative example with three stored fundamental memories of length $N=16$. These are,
\begin{align}
\bar{\xi^1}:&(1, 1, 1, 1, 1, 1, 1, 1, 1, 1, 1, 1, 1, 1, 1, 1),\nonumber\\
\bar{\xi^2}:&(1, 1, 1, 1, 1, 1, 1, 1, -1, -1, -1, -1, -1, -1, -1, -1),\nonumber\\
\bar{\xi^3}:&(1, 1, 1, 1, -1, -1, -1, -1, -1, -1, -1, -1, 1, 1, 1, 1),\nonumber
\end{align}
which are stored using the Hebbian rule (\ref{hebbrule}) in a network of $16\times 16$ fully connected interacting qubits. The DW2 provided software tool is used to find an embedding onto the native "chimera" graph on the chip. This requires only $133$ physical qubits for embedding. Each of the $16$ logical qubits are encoded as ferromagnetic chains of physical qubits with the largest encoded qubit being a chain of $11$ physical qubits and the smallest of size $5$. The maximum value of any coupling $|J_{ij}|=W_{ij}$ is $(3/2^4)$ with the minimum being $(1/2^4)$.

Note that these fundamental memories are mutually orthogonal, i.e. $\sum_{i}\xi^\mu_i\xi^\nu_i=\delta_{\mu,\nu}$, since their pairwise Hamming distances are all equal to $N/2=8$. The probe input vector we use is, 
\begin{align}
\bar{\chi}=\{-1, 1, 1, 1, -1, -1, -1, -1, -1, -1, -1, -1, 1, 1, 1, -1\}
\label{probe1}
\end{align}
whose Hamming distances from the fundamental memories are $d^1_\chi=10,~d^2_\chi=8,~d^3_\chi=2$. The energies of the fundamental memories $\bar{\xi^1},\bar{\xi^2},\bar{\xi^3}$ and input memory $\bar{\chi}$ w.r.t. the final problem Hamiltonian, $\hat{H}_P$, using $\bar{\chi}$ from Expression (\ref{probe1}) to determine the probe Hamiltonian part, $\hat{H}_{\text{probe}}$ in Eq.~(\ref{eq:probemem}), are shown in Fig.~(\ref{fig:example-energy}). We expect the bound on the maximum field strength (\ref{hbound}) to be $h^\mu_{\chi,\text{max}}=.75$ obtained for $\mu=3$ (closest memory in Hamming distance), but test the success probability of recall using quantum annealing at increasing values of $h$ starting from $h=2^{-5}$ to, well beyond $h^\mu_{\chi,\text{max}}$, upto $h=1.2$ in linear steps of $2^{-5}$. At each value of $h$ we anneal a 100 times and the number of times the machine returns the closest memory $\bar{\xi^3}$ - expressed as a percentage - gives us the success probability $P_{\text{success}}$, Fig.~(\ref{fig:example}). We find that the annealing success probability is very close to unity, and can be essentially made 1 if one imposes a majority vote on the percentage, i.e. a percentage value greater than 50~\% is understood as success probability of 1 for that particular memory, for most of the allowed region except for very small $h$ values.  To understand why this may be so we consider the different sources of error on DW2 in the next paragraph.

How close to perfect recall this empirically determined quantity $P_{\text{success}}$ is, depends on several factors \cite{santramax2sat,QAD}. First and foremost, DW2 operates at a non-zero, albeit small, temperature of $T\simeq 15 mK$. This means that the quantum states representing two bit strings at a Hamming distance of $d\leq k_BT/h$, with $k_B$ - the Boltzmann's constant, from each other should be considered as degenerate w.r.t. the probe Hamiltonian $\hat{H}_{\text{probe}}$. This counts as thermal error \cite{thermal-error} which is the hardest to mitigate. Secondly, the encoding of logical qubits into ferromagnetic chains of physical qubits, the longer the worse, introduces errors at the emdedding stage - encoding error - that may be reduced by adopting embedding algorithms that minimize qubit chain lengths. Next, the physical implementation of the flux qubits favors an individual spin to align in one direction compared to the opposite direction which introduces the so-called gauge error - which may be suppressed via gauge-averaging \cite{squid-qubit,QAD}. For the specific class of recall tasks in AMMs, gauge-error implies that the same pair, of fundamental memory set and probe vector, may have a different success probability if they are encoded with the signs of all their spins flipped. Finally, short annealing times can also cause diabatic errors \cite{thermal-error} that causes higher energy eigenstates to be populated - which we have tried to reduce in our own experiments by using the maximum possible annealing time on the machine, $T_{\text{anneal}}=20,000~ \mu S$, in each run in order to ensure that such transitions are minimized. However, we have not analyzed the energy gap $\Delta$ of our problems to determine whether $T_{\text{anneal}}>>\Delta^{-1}$.

For the small $h\leq .15$ region in Fig.~(\ref{fig:example}), which although is well within the bound $h^\mu_{\chi,\text{max}}=.75$ given by Ineq.~(\ref{hbound}), we observe that the annealing success probability is severely degraded. This is caused by, we suspect, a combination of one or more factors discussed in the previous paragraph. However, the strongest reason might be the thermal noise that dominates at small $h$ values. Note that the actual physical energy that the field strength $h$ represents is obtained by multiplying it with $B(t)$ appearing in the time dependent annealing Hamiltonian (\ref{eq:timedepham}). The maximum value of $B(t)$ is $\simeq 30~GHz$ at the end of the annealing process starting from close to zero at $t=0$. An order of magnitude calculation shows that $hB(t)$ with $h=.15$ is of the same order of magnitude as $k_BT$ with $T=15~mK$ - for at least half the annealing time. The reduced annealing success probability in this region of $h$ might thus be attributed to thermally caused leakage of population to other energy eigenstates.

\section{Discussion and conclusion}
\label{discconc}
We have shown that using quantum annealing for recall tasks in Associative memory models can lead to an exponential capacity for storage and this scheme works with probability 1 for sets of randomly chosen memories in the large network size limit. The positive exponents for capacity and that for the scheme success probability have to sum to a decreasing function of the radius of the attraction basins - hence the tradeoff between the radius and the capacity and scheme success probability. Implementation of our scheme on a physical quantum annealing device may suffer thermal, encoding, gauge and diabatic  errors that can lead to imperfect recall even when all theoretical conditions for successful recall are met. The effective experimental success probability is determined empirically and depends on several factors of the physical implementation.

QAR-AMM should work for every learning rule where the memory Hamiltonian is degenerate on all fundamental memory vectors. Even so, one needs to consider certain inherent theoretical limitations that we now point out. The energy degeneracy of the stored memories is lifted by an amount proportional to their Hamming distance from the input memory times the uniform field strength $h$. The maximum field strength value such that the system does not get overbiased is inversely proportional to the size of the problem, i.e., $h^\mu_{\xi,\text{max}}\sim 1/N$ \cite{neigovzen:2008}. This automatically sets the upper bound for the adiabatic energy gap of the problem because for two fundamental memories differing by a Hamming distance of $d$ - the energy difference w.r.t. the problem Hamiltonian is proportional to $d\times h^\mu_{\xi,\text{max}}$. This means that, at least, for storing a linear number of memories one can expect an efficient adiabatic (annealing) run time, $T_{\text{anneal}}=O(N^\alpha),~\alpha=\text{small positive integer}$, for the recall task. However, as the number of stored memories approaches, the theoretically achievable, exponential limit - the number of stored memories at the same Hamming distance from the input can grow exponentially - the number of possible stored memories at any distance $d$ is given by the binomial coefficient $\binom{N}{d}$. To break the degeneracy of these equidistant memories one can choose, instead of a uniform field $h$, a position dependent non-uniform field $h_i,~i=1,2,...,N$. However, for any finite range of $h_i$-values, i.e. $\delta h=\text{max}_i(h_i)-\text{min}_i(h_i)<\infty$, the degeneracy would be broken by an amount proportional to $\delta h/\binom{N}{d}$. This means that the adiabatic energy gap in the exponential storage limit would close, as an inverse exponential, making recall tasks inefficient - time complexity wise. Nevertheless, even a polynomial storage capacity, with a concomitant polynomial QAR-AMM run-time, is a significant improvement (see \cite{class-exp-mem} and references therein) for the case of completely general Hopfield networks considered here.

Going forward, we would like to explore equivalent recall schemes for forgetful learning rules that favor recently added fundamental memories in the learning set to the ones before \cite{storkeyffl}. There one would like to understand the minimal requirements on the additional types of terms in the problem Hamiltonian those recall schemes would need. Further, we note that the recall process may be considered as an adiabatic quantum error correction operation - a fundamental memory may be understood as a codeword and its basin of attraction as the codespace \cite{nielsen2000quantum}. Each input memory within the basin of attraction corresponds to a distinct error. The conditional error-correction operation is the unitary obtained as a result of the evolution under the time-dependent annealing Hamiltonian which depends on the input state. Clearly, the scheme of QAR-AMM does not \emph{detect} errors but only corrects them - but in doing so it greatly enhances the capacity of the classical Hopfield network models. This is yet another instance of a hybrid protocol where partitioning of a computational job into quantum and classical subtasks leads to distinct advantages. However, the question of optimality of such partitioning is still open \cite{Zick2015}. Finally, we comment that the QAR-AMM scheme also requires classical pre-processing to minor-embed the problem graph on the Dwave annealing architecture - a step that may be obviated through the use of fully connected quantum annealers as recently proposed in \cite{qa-zoller}.

\emph{Acknowledgements}- SS thanks Tameem Albash, Daniel Lidar, Travis Humble and Shashank Sethi for valuable feedback. OS thanks Adam Douglass and Murray Thom for insightful feedback.

\bibliographystyle{apsrev4-1}
\bibliography{refs-amm1}
\appendix
\section{Bound on field strength}
\label{hboundmax}
We require the energies of any state that is not a fundamental memory to be greater than that of the latter w.r.t. the Hamiltonian (\ref{eq:probemem}):
\begin{align}
&-\braket{\xi^\mu|W/2|\xi^\mu}-h\sum_i\chi_i\xi^\mu_i<-\braket{\chi|W/2|\chi}-h\sum_i\chi_i\chi_i\nonumber\\
&-\braket{\xi^\mu|W/2|\xi^\mu}-h(n-2d^\mu_\chi)<-\braket{\chi|W/2|\chi}-hn\nonumber\\
\end{align}
For the case of orthogonal fundamental memory vectors their energy is given by,
\begin{align}
\braket{\xi^\mu|W/2|\xi^\mu}&=\frac{1}{2}\sum_{i,j}\xi^{\mu}_iW_{ij}\xi^{\mu}_j\nonumber\\
&=\frac{1}{2N}\{\sum_{\nu}\sum_{ij}\xi^{\mu}_i(\xi^{\nu}_i\xi^{\nu}_j)\xi^{\mu}_j-p\sum_{ij}\delta_{ij}\xi^{\mu}_i\xi^{\mu}_j\}\nonumber\\
&=\frac{1}{2N}\{\sum_{\nu}(\sum_{i}\xi^{\mu}_i\xi^{\nu}_i)(\sum_{j}\xi^{\mu}_j\xi^{\nu}_j)-p\sum_{i}\xi^{\mu}_i\xi^{\mu}_i\}\nonumber\\
&=\frac{1}{2N}\{\sum_\nu (N\delta_{\mu,\nu})(N\delta_{\mu,\nu})-pN\}\nonumber\\
&=(N-p)/2,
\end{align}
whereas for any arbitary input vector $\bar{\chi}$,
\begin{align}
\braket{\chi|W/2|\chi}&=\frac{1}{2N}\{\sum_{\nu}\sum_{ij}\chi_i(\xi^{\nu}_i\xi^{\nu}_j)\chi_j-p\sum_{ij}\delta_{ij}\chi_i\chi_j\}\nonumber\\
&=\frac{1}{2N}\{\sum_{\nu}(\sum_{i}\chi_i\xi^{\nu}_i)(\sum_{j}\chi_j\xi^{\nu}_j)-p\sum_{i}\chi_i\chi_i\}\nonumber\\
&=\frac{1}{2N}\{\sum_{\nu}(\sum_{i}\chi_i\xi^{\nu}_i)(\sum_{j}\chi_j\xi^{\nu}_j)-p\sum_{i}\chi_i\chi_i\}\nonumber\\
&=\frac{1}{2N}\{\sum_{\nu=1}^p (N-2d^{\nu}_\chi)^2-pN\}\nonumber\\
&=\frac{1}{2}\{\frac{1}{N}\sum_{\nu=1}^p(N-2d^\nu_\chi)^2-p\},
\end{align}
where we have used $\sum_{i}\chi_i\xi^{\nu}_i=(N-2d^\nu_\chi)$.
\section{Annealing Schedule.}
\begin{figure}
\centering
    \includegraphics[width=.9\columnwidth,height=4.5cm]{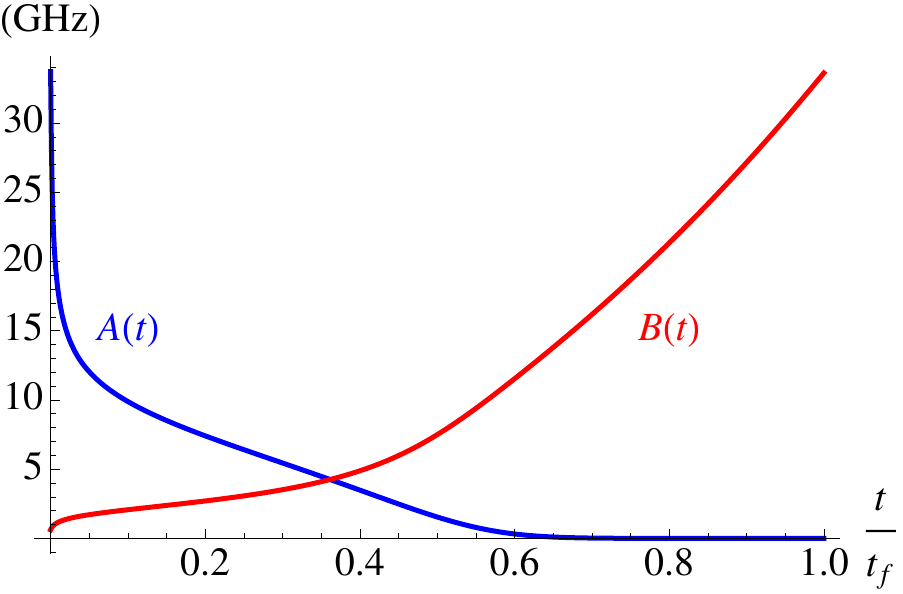}
    \caption{(color online) Temporal evolution of the classical control functions $A(t),~B(t)$ in the time-dependent annealing Hamiltonian $H(t)$ in Eq.~(\ref{eq:timedepham}).}
    \label{fig:annsch}
\end{figure}

Fig.~(\ref{fig:annsch}) shows the hard-coded classical controls $A(t),~B(t)$ evolving as a function of the scaled annealing time $t/T_{\text{anneal}}$ where $t$ is the physical time lapsed during the annealing process and $T_{\text{anneal}}$ is the user-set length of the annealing process. $T_{\text{anneal}}$ can be set at any integer value between a minimum of $20~\mu s$ to a maximum of $20,000~\mu s$.

\end{document}